\documentclass[runningheads,a4paper]{llncs}

\usepackage[pdftex]{graphicx}

\usepackage[english]{babel}

\usepackage[cmex10]{amsmath}
\usepackage{amssymb}

\usepackage{url}

\newcommand{\hairspace}{\hspace{1pt}}
\newcommand{\eg}{\mbox{e.\hairspace{}g.,} } 
\newcommand{\ie}{\mbox{i.\hairspace{}e.} }

\newcommand{\etal}{\mbox{et~al.}\ }

\usepackage{IEEEtrantools}
	
\usepackage{caption}
\DeclareCaptionLabelFormat{cornyTop}{Fig.~\thefigure #2 (top):}
\DeclareCaptionLabelFormat{cornyLeft}{Fig.~\thefigure #2 (left):}
\usepackage{subcaption}

\usepackage{endnotescorny}

\def\notesname{Definitions, Lemmata, and Theorems}
\renewcommand{\theendnote}{[{\roman{endnote}}]} 
\usepackage[symbol,perpage]{footmisc}

\usepackage{booktabs}
\usepackage{array}
\usepackage{multirow}

\usepackage{wrapfig}

\usepackage{floatrow}
\newfloatcommand{capbtabbox}{table}[][\FBwidth]

\usepackage{pifont}
\newcommand{\cmark}{\ding{51}}%
\newcommand{\xmark}{\ding{55}}%

\usepackage{expdlist}

\usepackage[pdftex,hyperfootnotes=false,pdfpagelabels,bookmarks,linktoc=section]{hyperref}
\hypersetup{
    colorlinks=true,
    linktocpage=true,
    breaklinks=true,
    bookmarksnumbered,
    bookmarksopen=true,
    bookmarksopenlevel=1,
    pdfhighlight=/O,
    pdfpagemode=UseOutlines,
    urlcolor=black,
    linkcolor=black,
    citecolor=black,
    pdfauthor={Cornelius Diekmann}
}

\newcommand{\studyTotalEntries}{15}

\newcommand{\studyNovicTopoValid}{12.0/10.6/5.7}
\newcommand{\studyNovicTopoInvalid}{4.0/6.6/4.9}
\newcommand{\studyNovicTopoMissing}{11.0/11.2/6.2}

\newcommand{\studyInterTopoValid}{14.0/14.0/1.4}
\newcommand{\studyInterTopoInvalid}{1.0/1.6/1.9}
\newcommand{\studyInterTopoMissing}{7.0/7.4/1.0}

\newcommand{\studyExperTopoValid}{16.0/15.8/1.7}
\newcommand{\studyExperTopoInvalid}{1.0/3.2/4.4}
\newcommand{\studyExperTopoMissing}{5.0/5.6/2.1}

\newcommand{\evalmodel}[0]{m}
\newcommand{\nP}[0]{P}

\newcommand{\tool}{\texttt{topoS}}

\begin{document}
\newcounter{proofobligationcounter}
\newtheorem{proofobligation}[proofobligationcounter]{PO}


\title{Verifying Security Policies using Host Attributes}

\author{Cornelius Diekmann\inst{1} \and Stephan-A. Posselt\inst{1} \and Heiko Niedermayer\inst{1} \and Holger Kinkelin\inst{1} \and Oliver Hanka\inst{2} \and Georg Carle\inst{1}}

\authorrunning{C. Diekmann \etal}

\institute{Technische Universit{\"a}t M{\"u}nchen \qquad \email{surname@net.in.tum.de}
\and Airbus Group Innovations \qquad \email{first\_name.surname@eads.net}}

\maketitle

\begin{abstract}
For the formal verification of a network security policy, it is crucial to express the verification goals. 
These formal goals, called security invariants, should be easy to express for the end user. 
Focusing on access control and information flow security strategies, this work discovers and proves universal insights about security invariants. 
This enables secure and convenient auto-completion of host attribute configurations.
We demonstrate our results in a civil aviation scenario. 
All results are machine-verified with the Isabelle/HOL theorem prover. 
\end{abstract}

\section{Introduction}
A distributed system, from a networking point of view, is essentially a set of interconnected hosts. 
Its connectivity structure comprises an important aspect of its overall attack surface, which can be dramatically decreased by giving each host only the necessary access rights. 
Hence, it is common to protect networks using firewalls and other forms of enforcing network level access policies.
However, raw sets of such policy rules \eg firewall rules, ACLs, or access control matrices, scale quadratically with the number of hosts and ``controlling complexity is a core problem in information security''~\cite{guttman05rigorous}. 
A case study, conducted in this paper, reveals that even a policy with only 10 entities may cause difficulties for experienced administrators. 
Expressive policy languages can help to reduce the complexity. 
However, the question whether a policy fulfills certain security invariants and how to express these often remains. 

\begin{figure}[!h]
  \centering
  		\includegraphics[width=0.99\linewidth]{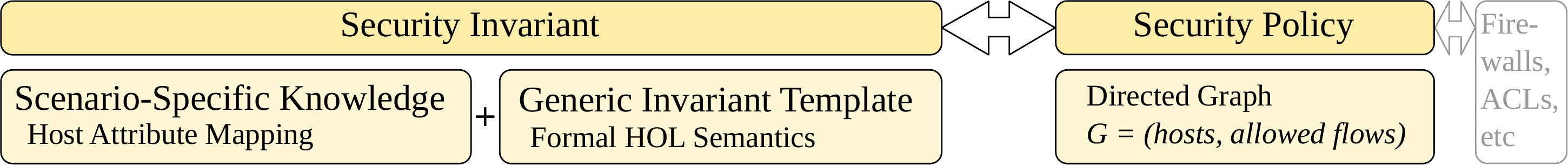}
\vskip-4pt%
  \caption{Formal objects: Security invariant and security policy.}%
  \label{fig:intro:formalobjects}%
\end{figure}

\noindent%
Using an attribute-based~\cite{abac2005} approach, we model simple, static, positive security policies with expressive, Mandatory Access Control (MAC) security invariants. 
The formal objects, illustrated in Fig.~\ref{fig:intro:formalobjects}, are carefully constructed for their use-case. 
The policy is simply a graph, which can for example be extracted from or translated to firewall rules. 
The security invariants are split into the formal semantics, accessible to formal analysis, and scenario-specific knowledge, easily configurable by the end user.  
This model landscape enables verification of security policies. 
Primarily, we contribute the following universal insights for constructing security invariants. 
\noindent\begin{enumerate}
	\item Both provably \emph{secure} and \emph{permissive} default values for host attributes can be found. This auto completion decreases the user's configuration effort. 
	\item The security strategy, information flow or access control, determines whether a security violation occurs at the sender's or at the receiver's side. 
	\item A violated invariant can always be repaired by tightening the policy \emph{if and only if} the invariant holds for the deny-all policy. 
\end{enumerate}
We formally introduce the underlying model in Sect.~\ref{sec:formalmodel}.
Then we present three examples of security invariant templates in Sect.~\ref{sec:model-library} and conduct a formal analysis in Sect.~\ref{sec:analysis}. 
Our implementation and a case study are presented in Sections~\ref{sec:impl} and~\ref{sec:case-study}. 
Related work is described in Sect.~\ref{sec:relatedwork}.
We conclude in Sect.~\ref{sec:conclusion}. 

\section{Formal Model}
\label{sec:formalmodel}
We formalized all our theory in the Isabelle/HOL theorem prover~\cite{isabelle2013}.
To stay focused, we omit all proofs in this document but point the reader to the complete formal proofs by roman reference marks. 
For example, when the paper states `foo\endnotemark[4]', the machine-verified proof for the claim `foo' can be found by following the corresponding endnote. 
Note that standards such as Common Criteria~\cite{cc2012p3} require formal verification for their highest \emph{Evaluation Assurance Level} (EAL7) and the Isabelle/HOL theorem prover is suitable for this purpose~\cite[\S{}A.5]{cc2012p3}. 
Therefore, our approach is not only suitable for verification, but also a first step towards certification.

Retaining network terminology, we will use the term \emph{host} for any entity which may appear in a policy\footnote{In contrast to common policy terminology, we do not differentiate between subjects and targets (objects) as they are usually indistinguishable on the network layer and a host may act as both. }, \eg collections of IP addresses, names, or even roles. 
A \emph{security policy} is ``a specific statement of what is and is not allowed''~\cite{bishop2003compsec}. 
Narrowing its scope to network level access control, a security policy is a set of rules which state the allowed communication relationships between hosts. 
It can be represented as a directed graph. 

\begin{definition}[Security Policy]
\label{def:securitypolicy}
	A security policy is a directed graph $G = (V,\, E)$, where the hosts $V$ are a set of type \mbox{$\mathcal{V}$} and the allowed flows $E$ are a set of type \mbox{$\mathcal{V}{\times}\mathcal{V}$}. 
	The type of $G$ is abbreviated by $\mathcal{G} = (\mathcal{V}\;set) \times ((\mathcal{V}{\times}\mathcal{V})\;set)$. 
\end{definition}

\noindent
A policy defines rules (\emph{``how?''}). It does not justify the intention behind these rules (\emph{``why?''}). 
To reflect the \emph{why?}-question, we note that depending on a concrete scenario, hosts may have varying security-relevant attributes. 
We model a host attribute of arbitrary type $\Psi$ and establish a total mapping from the hosts $V$ to their scenario-specific attribute. 
Security invariants can be constructed by combining a \emph{host mapping} with a \emph{security invariant template}. 
Latter two are defined together because the same $\Psi$ is needed for a related {host mapping} and {security invariant template}. 
Different $\Psi$ may appear across several security invariants. 

\begin{definition}[Host Mapping and Security Invariant Template]
\label{def:securityinvarianttemplate}
For scenario-specific attributes of type $\Psi$, a host mapping $\nP$ is a total function which maps a host  to an attribute. 
$\nP$ is of type $\mathcal{V} \Rightarrow \Psi$. 

A security invariant template $\evalmodel$ is a predicate\footnote{a predicate is a total, Boolean-valued function.} $\evalmodel(\mathcal{G}, (\mathcal{V} \Rightarrow \Psi))$, defining the formal semantics of a security invariant. 
Its first argument is a security policy, its second argument a host attribute mapping. 
The predicate \mbox{${\evalmodel}(G, \nP)$} returns true iff the security policy $G$ fulfills the security invariant specified by $\evalmodel$ and $\nP$. 
\end{definition}

\begin{example}
\label{example:blp}
Label-based information flow security can be modeled with a simplified version of the Bell LaPadula model~\cite{blphistory,bell1973secure2}. 
Labels, more precisely \emph{security clearances}, are host attributes $\Psi = \left\lbrace\mathit{unclassified},\, \mathit{confidential},\, \mathit{secret},\, \mathit{top\-secret}\right\rbrace$. 
The Bell LaPadula's no read-up and no write-down rules can be summarized by requiring that the security clearance of a receive\mbox{r $r$} should be greater-equal than the security clearance of the sende\mbox{r $s$}, for all $(s,r) \in E$. 
With a total order `$\leq$' on $\Psi$, the security invariant template can be defined as $\evalmodel((V,\,E),\, \nP) \equiv {\forall (s,r) \in E.}\ \nP(s) \leq \nP(r)$. 

Let the scenario-specific knowledge be that database $\mathit{db_1} \in V$ is $\mathit{confidential}$ and all other hosts are $\mathit{unclassified}$. 
Using lambda calculus, the total function $\nP$ can be defined as $(\lambda h. \ \mathbf{if}\ h = \mathit{db_1} \ \mathbf{then} \ \mathit{confidential} \ \mathbf{else} \ \mathit{unclassified})$. 
Hence $\nP(\mathit{db}_1) = \mathit{confidential}$. 
For any policy $G$, the predicate $\evalmodel(G,\, \nP)$ holds if $\mathit{db}_1$ does not leak confidential information (\ie there is no non-reflexive outgoing edge from $\mathit{db}_1$). 
\end{example}

\noindent
Security invariants formalize security goals. 
A template contributes the formal semantics. 
A host mapping contains the scenario-specific knowledge. 
This makes the scenario-independent semantics available for formal reasoning by treating $\nP$ and $G$ as unknowns. 
Even reasoning with arbitrary security invariants is possible by additionally treating $\evalmodel$ as unknown. 

With this modeling approach, the end user needs not to be bothered with the formalization of $\evalmodel$, but only needs to specify $G$ and $\nP$. 
In the course of this paper, we present a convenient method for specifying $\nP$. 
\subsubsection{Security Strategies and Monotonicity.}
In IT security, one distinguishes between two main classes of security strategies: \emph{Access Control Strategies} (ACS) and \emph{Information Flow Strategies} (IFS) \cite[\S\hairspace6.1.4]{eckert2013}.
An IFS focuses on confidentiality and an ACS on integrity or controlled access. 
We require that $\evalmodel$ is in one of these classes\footnote{By limiting $\evalmodel$ to IFS or ACS, we emphasize that availability is not in the scope of this work. Availability requires reasoning on a lower abstraction level, for example, to incorporate network hardware failure. Availability invariants could be expressed similarly, but would require inverse monotonicity (see below). }.

The two security strategies have one thing in common: 
they prohibit illegal actions. 
From an integrity and confidentiality point of view, prohibiting more never has a negative side effect.  
Removing edges from the policy cannot create new accesses and hence cannot introduce new access control violations. 
Similarly, for an IFS, by statically prohibiting flows in the network, no new direct information leaks nor new side channels can be created. 
In brief, prohibiting more does not harm security. 
From this, it follows that if a policy $(V,\, E)$ fulfills its security invariant, for a stricter policy rule set $E' \subseteq E$, the policy $(V,\, E')$ must also fulfill the security invariant. 
We call this property \emph{monotonicity}.

\subsubsection{Composition of Security Invariants.}
Usually, there is more than one security invariant for a given scenario. 
However, composition and modularity is often a non-trivial problem. 
For example, access control lists that are individually secure can introduce security breaches under composition \cite{composable94}. 
Also, information flow security of individually secure processes, systems, and networks may be subverted by composition~\cite{restrictiveness}. 
This is known as the \emph{composition problem}~\cite{blphistory}. 

With the formalization in this paper, composability and modularity are enabled by design. 
For a fixed policy $G$ with $k$ security invariants, let $\evalmodel{}_i$ be the security invariant template and $\nP{}_i$ the host mapping, for $i \in \left\lbrace 1 \dots k \right\rbrace$. 
The predicate $\evalmodel_i(G,\, \nP_i)$ holds if and only if the security invariant $i$ holds for the policy $G$. 
With this modularity, composition of all security invariants is straightforward\endnote{all-security-requirements-fulfilled}: all security invariants must be fulfilled. 
The monotonicity guarantees that having more security invariants provides greater or equal security. 
\vskip-14pt 
\begin{IEEEeqnarray*}{c}
\evalmodel_1(G,\, \nP_1) \wedge \dots \wedge \evalmodel_k(G,\, \nP_k)
\end{IEEEeqnarray*}

\section{Examples of Security Invariant Templates}
\label{sec:model-library}
In this section, we present three examples of security invariant templates. 
Our implementation currently features more than ten templates and grows. 
All can be inspected in the published theory files. 
Common networking scenarios such as subnets, non-interference invariants, or access control lists are available. 
With the following templates, a larger case study is presented in Section~\ref{sec:case-study}.

\subsubsection{Simplified Bell LaPadula with Trust.}
A simplified version of the Bell LaPadula model is already outlined in Example~\ref{example:blp}. 
In this paragraph, we extend this model with a notion of trust by adding a Boolean flag \emph{trust} to the host attributes. 
For a host $v$, let $\nP(v).\mathit{sc}$ denote $v$'s security clearance and $\nP(v).\mathit{trust}$ if $v$ is trusted. 
A trusted host can receive information of any security clearance and may declassify it, \ie distribute the information with its own security clearance. 
For example, a trusted host is allowed to receive any information and with the $\mathit{unclassified}$ clearance, it is allowed to reveal it to anyone. 
The template is thus formalized as follows. 
\begin{IEEEeqnarray*}{ll}
  \evalmodel\bigl((V, E),\, \nP\bigr) \ \equiv \ \forall (s, r) \in E. \ \begin{cases} 
  	\mathit{True} & \mathbf{if} \ \ \nP(r).\mathit{trust} \\ 
  	\nP(s).\mathit{sc} \; \leq \; \nP(r).\mathit{sc} & \mathbf{otherwise}
  	\end{cases}
\end{IEEEeqnarray*}

\subsubsection{Domain Hierarchy.}
The domain hierarchy template mirrors hierarchical access control structures. 
\begin{figure*}[htbp]
  \centering
  \hspace*{\fill}%
  \begin{subfigure}[t]{0.64\textwidth}
       \captionsetup{width=0.99\textwidth}
       \centering
  		\includegraphics[width=0.99\linewidth]{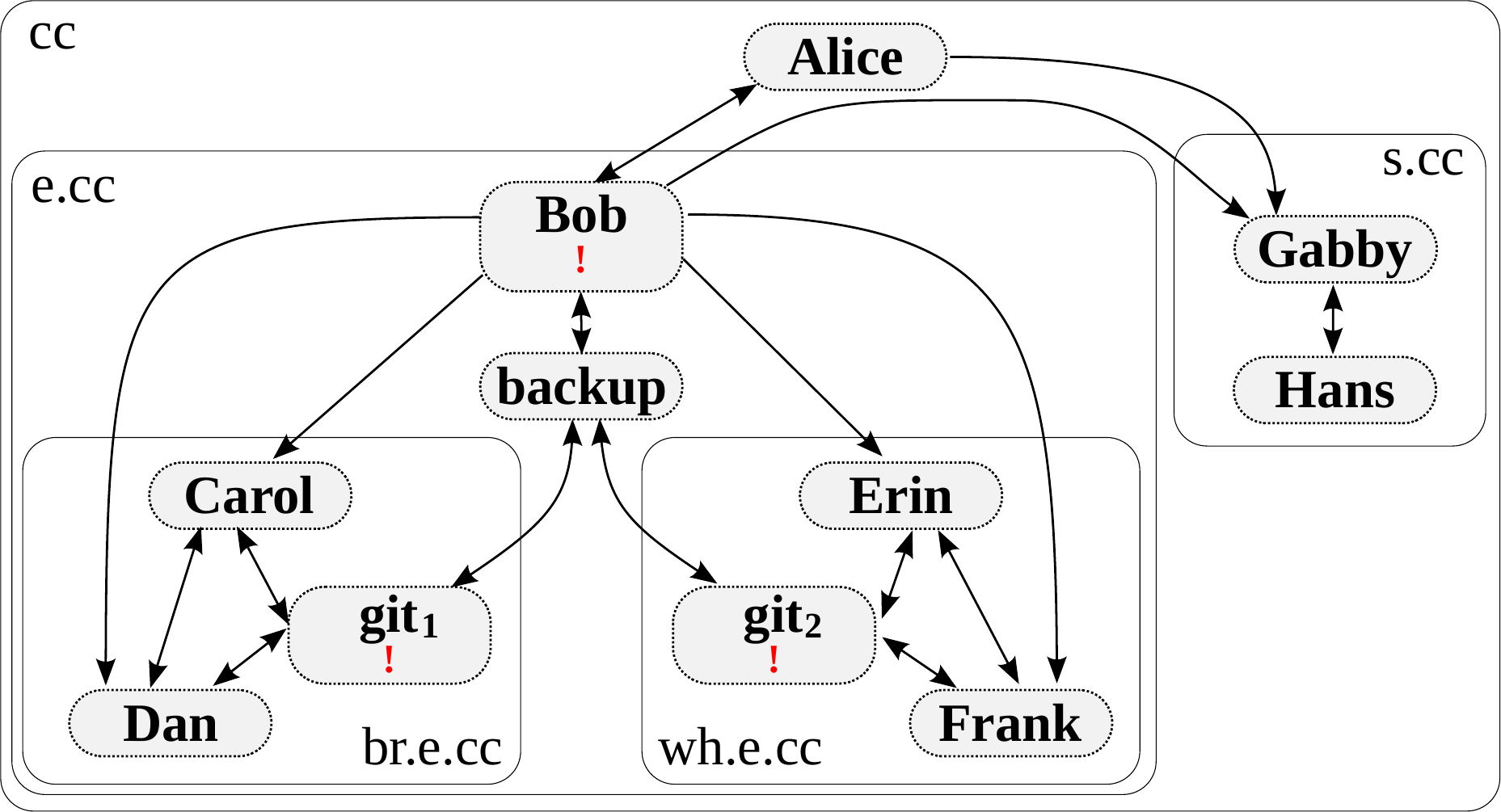}
  \end{subfigure}%
  \hspace*{\fill}%
  \begin{subfigure}[b]{0.34\textwidth}
       \captionsetup{width=0.85\textwidth,labelformat=cornyTop}
       \centering
  		\includegraphics[width=0.50\linewidth]{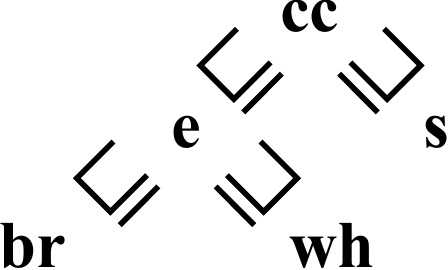}
  		\vspace*{2em}
  		\caption{Organizational structure of \textbf{cc}.}
  		\label{fig:domain_hierarchy_example_cc}
       \captionsetup{width=0.85\textwidth,labelformat=cornyLeft}
		\caption{Policy and host mapping of \textbf{cc}.}
		\label{fig:domhierarchynetworkcc}
  \end{subfigure}%
  \hspace*{\fill}%
\end{figure*}
%
\refstepcounter{figure}%
It is best introduced by example. 
The tiny car company (\textit{cc}) consists of the two sub-departments engineering (\textit{e}) and sales (\textit{s}). 
The engineering department itself consists of the brakes (\textit{br}) and the wheels (\textit{wh}) department. 
This tree-like organizational structure is illustrated in Fig.~\ref{fig:domain_hierarchy_example_cc}. 
We denote a position by the fully qualified domain name, 
\eg \textit{wh.e.cc} uniquely identifies the wheels department. 
Let `$\sqsubseteq$' denote the `\emph{is below or at the same hierarchy level}' relation, \eg $\mathit{wh}.e.\mathit{cc} \sqsubseteq \mathit{wh}.e.\mathit{cc}$, $\mathit{wh}.e.\mathit{cc} \sqsubseteq e.\mathit{cc}$, and $\mathit{wh}.e.\mathit{cc} \sqsubseteq \mathit{cc}$. 
However, $ \mathit{wh}.e.\mathit{cc} \not\sqsubseteq \mathit{br}.e.\mathit{cc}$ and $\mathit{br}.e.\mathit{cc} \not\sqsubseteq \mathit{wh}.e.\mathit{cc}$. 
The `$\sqsubseteq$' relation denotes a partial order\endnote{instantiation domainNameDept :: order}. 
The company's command structures are strictly hierarchical, \ie commands are either exchanged in the same department or travel from higher departments to their sub-departments. 
Formally, the receiver's level $\sqsubseteq$ sender's level. 
For a host $v$, let $\nP{}(v).\mathit{level}$ map to the fully qualified domain name of $v$'s department. 
For example in Fig.~\ref{fig:domhierarchynetworkcc}, $\nP{}(\mathit{Bob}).\mathit{level} = e.\mathit{cc}$. 

As in many real-world applications of a mathematical model, exceptions exist. 
Those are depicted by exclamation marks in Fig.~\ref{fig:domhierarchynetworkcc}. 
For example, Bob as head of engineering is in a trusted position. 
This means he can operate as if he were in the position of Alice. 
This implies that he can communicate on par with Alice, which also implies that he might send commands to the sales department. 
We model such exceptions by assigning each host a trust level. 
This trust level specifies up to which position in the hierarchy this host may act. 
For example, Bob in \textit{e.cc} with a trust level of $1$ can act as if he were in \textit{cc}, which means he has the same command power as Alice. 
Let $\nP{}(v).\mathit{trust}$ map to $v$'s trust level. 
We implement a function $\mathit{chop}(\mathit{level}: \mathit{DomainName},\  \mathit{trust}: \mathbb{N}) \Rightarrow \mathit{DomainName}$ which chops off $\mathit{trust}$ sub-domains from a domain name, \eg $\mathit{chop}(\mathit{br.e.cc},\ 1) = \mathit{e.cc}$. 
With this, the security invariant template can be formalized as follows. 
\vskip-16pt 
\begin{IEEEeqnarray*}{l}
\evalmodel\bigl((V, E),\, \nP\bigr) \ \equiv \ \forall (s, r) \in E. \ \ \nP{}(r).\mathit{level} \; \sqsubseteq \; \mathit{chop}\bigl(\nP{}(s).\mathit{level},\ \nP{}(s).\mathit{trust}\bigr)
\end{IEEEeqnarray*}

\subsubsection{Security Gateway.}
Hosts may belong to a certain domain. 
Sometimes, a pattern where intra-domain communication between domain members must be approved by a central instance is required. 
As an example, let several virtual machines belong to the same domain and a secure hypervisor manage intra-domain communication. 
As another example, inter-device communication of slave devices in the same domain is controlled by a central master device. 
We call such a central instance `security gateway' and present a template for this architecture. 
Four host roles are distinguished: A security gateway ($\mathit{sgw}$), a security gateway accessible from the outside ($\mathit{sgwa}$), a domain member ($\mathit{memb}$), and a default value that reflects `none of these roles' ($\mathit{default}$). 
The following table implements the access control restrictions. 
The role of the sender (snd), role of the receiver (rcv), the result (rslt), and an explanation are given. 

\begin{table}[h!tbp]
\centering
\begin{tabular}[0.99\linewidth]{ @{}l@{ \ }l@{ }|c@{ \ }p{28em} }
	\toprule
	snd & rcv & rslt & explanation \\
	\midrule
 $\mathit{sgw}$ & * & \cmark & Can send to the world. \\
 $\mathit{sgwa}$ & * & \cmark & --- \grqq ---\\
 $\mathit{memb}$ & $\mathit{sgw}$ & \cmark & Can contact its security gateway. \\
 $\mathit{memb}$ & $\mathit{sgwa}$ & \cmark & --- \grqq ---\\
 $\mathit{memb}$ & $\mathit{memb}$ & \xmark & Must not communicate directly. May communicate via $\mathit{sgw}(a)$. \\
 $\mathit{memb}$ & $\mathit{default}$ & \cmark & No restrictions for direct access to outside world. Outgoing accesses are not within the invariant's scope. \\
 $\mathit{default}$ & $\mathit{sgw}$ & \xmark & Not accessible from outside. \\
 $\mathit{default}$ & $\mathit{sgwa}$ & \cmark & Accessible from outside. \\
 $\mathit{default}$ & $\mathit{memb}$ & \xmark & Protected from outside world. \\
 $\mathit{default}$ & $\mathit{default}$ & \cmark & No restrictions. \\
 \bottomrule%
\end{tabular}%
\end{table}

This template is minimalistic in that it only restricts accesses to members (from other members or the outside world), whereas accesses from members to the outside world are unrestricted. 
It can be implemented by a simple table lookup. 
In-host communication is allowed by adding $s \neq r$. 
\vskip-18pt
\begin{IEEEeqnarray*}{l}
\evalmodel\bigl((V, E),\, \nP\bigr) \ \equiv \ \forall (s, r) \in E,\ s \neq r. \ \ \mathtt{table}\bigl(\nP{}(s),\ \nP{}(r)\bigr)
\end{IEEEeqnarray*}

\section{Generic Semantic Analysis of Security Invariants}
\label{sec:analysis}
\subsubsection{Offending Flows.}
Since $\evalmodel$ is monotonic, if an IFS or ACS security invariant is violated, there must be some flows in $G$ that are responsible for the violation. 
By removing them, the security invariant should be fulfilled (if possible). 
We call a minimal set of such flows the \emph{offending flows}. 
Minimality is expressed by requiring that every single flow in the offending flows bears responsibility for the security invariant's violation. 

\begin{definition}[Set of Offending Flows]
\label{def:set_offending_flows_def}
\begin{IEEEeqnarray*}{l}
\mathit{set\_offending\_flows}\bigl(G,\, \nP\bigr) = 
\bigl\lbrace F \subseteq E \ \vert \ \neg\, \evalmodel\bigl(G,\, \nP\bigr) \ \wedge \ \evalmodel\bigl((V,\ E \setminus F),\, \nP\bigr) \ \wedge \ \\ %
\IEEEeqnarraymulticol{1}{r}{\forall (s,r) \in F.\ \neg\, \evalmodel\bigl(\left(V,\, \left(E \setminus F\right) \cup \left\lbrace \left( s,r \right) \right\rbrace \right),\,  \nP\bigr) \bigr\rbrace}
\end{IEEEeqnarray*}
\end{definition}
\begin{example}
The definition does not require that the offending flows are uniquely defined. 
This is reflected in its type since it is a set of sets. 
For example, for $G = (\lbrace v_1, v_2, v_3 \rbrace, \lbrace (v_1, v_2), (v_2, v_3) \rbrace)$ and a security invariant that $v_1$ must not transitively access $v_3$, the invariant is violated: $v_2$ could forward requests. 
The set of offending flows is $\left\lbrace  \lbrace (v_1, v_2) \rbrace, \lbrace (v_2, v_3) \rbrace \right\rbrace$.
This ambiguity tells the end user that there are multiple options to fix a violated security invariant. 
The policy can be tightened by prohibiting one of the offending flows, $\eg \lbrace (v_1, v_2) \rbrace$. 
\end{example}
\noindent
If $\evalmodel(G,\,\nP)$ holds, the set of offending flows is always empty\endnote{validmodel-imp-no-offending}. 
Also, for every element in the set of offending flows, it is guaranteed that prohibiting these flows leads to a fulfilled security invariant\endnote{remove-offending-flows-imp-model-valid}. 
It is not guaranteed that the set of offending flows is always non-empty for a violated security invariant. 
Depending on $\evalmodel$, it may be possible that no set of flows satisfies Def.~\ref{def:set_offending_flows_def}. 
However, Theorem~\ref{thm:no-edges-validity} proves\endnote{valid-empty-edges-iff-exists-offending-flows} an important insight: a violated invariant can always be repaired by tightening the policy if and only if the invariant holds for the deny-all policy. 
\begin{theorem}[No Edges Validity]
\label{thm:no-edges-validity}
For $\evalmodel$ monotonic, arbitrary \mbox{$V,\, E$, and $\nP,$} let $G = (V,\, E)$ and $G_\mathit{deny\mbox{\textit{\small-}}all} = (V, \emptyset)$. If $\;\neg\, \evalmodel(G, \nP)$ then%
\vskip-14pt
\begin{IEEEeqnarray*}{l}
\evalmodel\bigl(G_\mathit{deny\mbox{\textit{\small-}}all},\, \nP\bigr) \longleftrightarrow \mathit{set\_offending\_flows}(G,\, \nP) \neq \emptyset
\end{IEEEeqnarray*}
\end{theorem}

\noindent
We demand that all security invariants fulfill $\evalmodel\bigl(G_\mathit{deny\mbox{\textit{\small-}}all},\, \nP\bigr)$. 
This means that violations are always fixable. 

We call a host responsible for a security violation the \emph{offending host}. 
Given one offending flow, the violation either happens at the sender's or the receiver's side. 
The following difference between ACS and IFS invariant can be observed. 
If $\evalmodel$ is an ACS, the host that initiated the request provokes the violation by violating an access control restriction. 
If $\evalmodel$ is an IFS, the information leak only occurs when the information reaches the unintended receiver. 
This distinction is essential as it renders the upcoming Def.~\ref{def:secure_default} and~\ref{def:unique_default} provable. 

\begin{definition}[Offending Hosts]
\label{def:offenders} For \mbox{$F \in \mathit{set\_offending\_flows}(G, \nP)$}
\vskip-14pt
	\begin{IEEEeqnarray*}{lcll}
	\mathit{offenders}(F) & \ = \ & \begin{cases} \left\lbrace \;s \ \vert \ (s, r) \in F \right\rbrace & \mathbf{if} \ \text{ACS}  \\
	\left\lbrace \;r \ \vert \ (s, r) \in F \right\rbrace  & \mathbf{if} \ \text{IFS}\end{cases}
	\end{IEEEeqnarray*}
\end{definition}

\subsubsection{Secure Auto Completion of Host Mappings.}
\label{sec:securedefault} Since $\nP$ is a \emph{total} function $\mathcal{V} \Rightarrow \Psi$, a host mapping for \emph{every} element of $\mathcal{V}$ must be provided. 
However, an end user might only specify the \emph{security-relevant} host attributes. 
Let $\nP_C \subseteq \mathcal{V} \times \Psi$ be a finite, possibly incomplete host attribute mapping specified by the end user. 
For some $\bot \in \Psi$, the total function $\nP$ can be constructed by $\nP(v) \equiv (\mathbf{if}\ {(v, \psi) \in \nP_C} \ \mathbf{then} \ \psi \ \mathbf{else} \ \bot)$.
Intuitively, if no host attribute is specified by the user, $\bot$ acts as a default attribute. 

Given the user specified all security-relevant attributes, we observe that the default attribute can never solve an existing security violation. 
Therefore, we conclude that for a given security invariant $\evalmodel$, a value $\bot$ can securely be used as a default attribute if it cannot mask potential security risks. 
In other words, a default attribute $\bot$ is secure w.r.t.\ the given information $\nP$ if for all \mbox{offenders $v$}, replacing $v$'s attribute\footnote{$\nP_{v \mapsto \bot} \equiv (\lambda x. \ \mathbf{if}\ x = v \ \mathbf{then} \ \bot \ \mathbf{else} \ \nP(x))$, an updated $\nP$ which returns $\bot$ for $v$} by $\bot$, denoted by $\nP_{v \mapsto \bot}$, has the same amount of security-relevant information as the original $\nP$.

\begin{definition}[Secure Default Attribute]
\label{def:secure_default}
A $\bot$ is a secure default attribute iff for a fixed $\evalmodel$ and for arbitrary $G$ and $\nP$ that cause a security violation, replacing the host attribute of any offenders by $\bot$ must guarantee that no security-relevant information is masked. 
\vskip-14pt
\begin{IEEEeqnarray*}{l}
\forall \; G \; \nP{}. \ \forall \; F \in \mathit{set\_offending\_flows}\bigl(G, \nP\bigr). 
\ \ \forall v \in \mathit{offenders}(F).\ \neg\, \evalmodel\bigl(G, \nP_{v \mapsto \bot}\bigr)
\end{IEEEeqnarray*}
\end{definition}

\begin{example}
In the simple Bell LaPadula model, an IFS, let us assume information is leaked. 
The predicate `information leaks' holds, no matter to which lower security clearance the information is leaked.
In general, if there is an illegal flow, it is from a higher security clearance at the sender to a lower security clearance at the receiver. 
Replacing the security clearance of the receiver with the lowest security clearance, the information about the security violation is always preserved. 
Thus, $\mathit{unclassified}$ is the secure default attribute\endnote{interpretation BLPbasic: NetworkModel}. 
In summary, if all classified hosts are labeled correctly, treating the rest as unclassified prevents information leakage. 
\end{example}

\noindent
To elaborate on Def.~\ref{def:secure_default}, it can be restated as follows. 
It focuses on the available security-relevant information in the case of a security violation. 
The attribute of an offending host $v$ bears no information, except for the fact that there is a violation. 
A secure default attribute $\bot$ cannot solve security violations. 
Hence $\nP(v)$ and $\bot$ are equal w.r.t.\ the security violation. 
Thus, $\nP$ and $\nP_{v \mapsto \bot}$ must be equal w.r.t.\ the information about the security violation.  
Requiring this property for all policies, all possible security violations, all possible choices of offending flows, and all candidates of offending hosts, this definition justifies that $\bot$ never hides a security problem.

\begin{example}
Definition~\ref{def:secure_default} can be specialized to the exemplary case in which a new host $x$ is added to a policy $G$ without updating the host mapping. 
Consulting an oracle, $x$'s real host attribute is $\nP(x) = \psi$. 
In reality, the oracle is not available and $x$ is mapped to $\bot$ because it is new and unknown. 
Let $x$ be an attacker. 
With the oracle's $\psi$-attribute, $x$  causes a security violation. 
We demand that the security violation is exposed even without the knowledge from the oracle. 
Definition~\ref{def:secure_default} satisfies this demand: if $x$ mapped by the oracle to $\psi$ causes a security violation, $x$ mapped to $\bot$ does not mask the security violation. 
\end{example}

\noindent
A `deny-all' default attribute is easily proven secure. 
Definition~\ref{def:secure_default} reads the following for this case: 
if an offender $v$ does something that violates $\evalmodel(G, \nP)$, then removing all of $v$'s rights ($\nP_{v \mapsto \mathit{deny\mbox{\textit{\small-}}all}}$), a violation must persist. 
Hence, designing whitelisting security invariant templates with a restrictive default attribute is simple. 
However, to add to the ease-of-use, more permissive default attributes are often desirable since they reduce the manual configuration effort. 
In particular, if a security invariant only concerns a subset of a policy's hosts, no restrictions should be imposed on the rest of the policy. 
This is also possible with Def.~\ref{def:secure_default}, but may require a comparably difficult proof.

\begin{example}
In Example~\ref{example:blp}, no matter how many hosts are added to the policy, it is sufficient to only specify that $\mathit{db_1}$ is $\mathit{confidential}$. 
This confidentiality is guaranteed while no restrictions are put on hosts that do not interact with $db_1$. 
\end{example}

\begin{definition}[Default Attribute Uniqueness]
\label{def:unique_default}
A default attribute $\bot$ is called unique iff it is secure (Def.~\ref{def:secure_default}) and there is no $\bot' \neq \bot$ s.t. $\bot'$ is secure. 
\end{definition}%
We demand that all security invariants fulfill Def.~\ref{def:unique_default}. 
This means that there is only one unique secure default attribute $\bot$. 
\begin{example}%
In the simple Bell LaPadula model, since the security clearances form a total order, the lowest security clearance is uniquely defined. 
\end{example}
\noindent
With the experience of proving Def.~\ref{def:secure_default} and \ref{def:unique_default} for default attributes for 18 invariant templates, the connection between offending host and security strategy was discovered. 
During our early research, we realized that a Boolean variable, fixed for $\evalmodel$, indicating the offending host was necessary to make Def.~\ref{def:secure_default} and \ref{def:unique_default} provable. 
A classification of the different invariants revealed the important connection.

\subsubsection{Default Attributes of Section~\ref{sec:model-library}'s Templates.}
In the Bell LaPadula with Trust template, the default attribute is $(\mathit{unclassified}, \mathit{untrusted})$\endnote{interpretation BLPtrusted: NetworkModel}. 
In the Domain Hierarchy, it is\endnote{interpretation DomainHierarchyNG: NetworkModel} a special value $\bot$ with a trust of zero and which is at the lowest point in the hierarchy, \ie $\forall\; l.\ \bot \sqsubseteq l$. 
Finally, it is worth mentioning that the $\sqsubseteq$-relation forms a lattice\endnote{instantiation domainName :: lattice}, which is a desirable structure for security classes~\cite{denning1976lattice}. 
In the Security Gateway, the default attribute is $\mathit{default}$\endnote{interpretation SecurityGatewayExtended-simplified: NetworkModel}.

All default attributes allow flows between each other. 
This greatly adds to the ease-of-use, since the scope of an invariant is limited only to the explicitly configured hosts. 
The unconcerned parts of a security policy are not negatively affected.

\subsubsection{Unique and Efficient Offending Flows.}
\label{sec:maxpolicmethodandoffending}
All security invariant templates presented in Section~\ref{sec:model-library} have a simple, common structure: a predicate is evaluated for all flows. 
Let $\Phi (\Psi, \Psi)$ be this predicate. 
Note that all invariants of this structure fulfill monotonicity\endnote{monotonicity-eval-model-mono}. 

Since Def.~\ref{def:set_offending_flows_def} is defined over all subsets, the naive computational complexity of is in \textbf{\textit{NP}}. 
This section shows that -- with knowledge about a concrete security invariant template $\evalmodel$ -- it can be computed in linear time. 
For $\Phi$-structured invariants, the offending flows are always uniquely defined and can be described intuitively\endnote{ENF-offending-set}\footnote{The same holds for templates with a structure similar to the Security Gateway\endnote{ENFnr-offending-set}}. 
\begin{theorem}[$\Phi$ Set of Offending Flows]
\label{thm:set_offending_flows_def}
If $\evalmodel$ is $\Phi$-structured $\evalmodel(G, \nP) \equiv \forall (s,r) \in E.\ \Phi(\nP(s),\, \nP(r))$, then
\vskip-12pt
	\begin{IEEEeqnarray*}{lcll}
	\mathit{set\_offending\_flows}\bigl(G,\, \nP\bigr) & \ = \ & \begin{cases} \lbrace\lbrace (s,r) \in E\ \vert \ \neg\;\Phi(\nP(s), \nP(r)) \rbrace\rbrace & \mathbf{if} \ \neg\;\evalmodel(G,\, \nP)  \\
	\emptyset  & \mathbf{if} \ \evalmodel(G,\, \nP)\end{cases}
	\end{IEEEeqnarray*}
\end{theorem}
\begin{example}
For the Bell LaPadula model, if no security violation exists the set of offending flows is $\emptyset$, else $\lbrace\lbrace (s,r) \in E\ \vert \ \nP(s) > \nP(r) \rbrace\rbrace$\endnote{BLP-offending-set}. 
\end{example}

\subsubsection*{Policy Construction.}
A policy that fulfills all security invariants can be constructed by removing all offending flows from the allow-all policy $G_{all} = (V, V{\times}V)$. 
This approach is sound\endnote{generate-valid-topology-sound} for arbitrary $\evalmodel$ and even complete\endnote{generate-valid-topology-max-topo} for $\Phi$-structured security invariant templates. 

\begin{example}
If completely contradictory security invariants are given, the resulting (maximum) policy is the deny-all policy $G_\mathit{deny\mbox{\textit{\small-}}all} = (V, \emptyset)$. 
\end{example}

\section{Implementation}
\label{sec:impl}
We built a tool called \tool{} with all the features presented in this paper. 
Its core reasoning logic consists of code generated by Isabelle/HOL. 
This guarantees the correctness of all results computed by \tool{}'s core~\cite{haftmann2010code}. 

\begin{figure}%
\begin{floatrow}%
\ffigbox{%
\centering%
  		\includegraphics[width=0.99\linewidth]{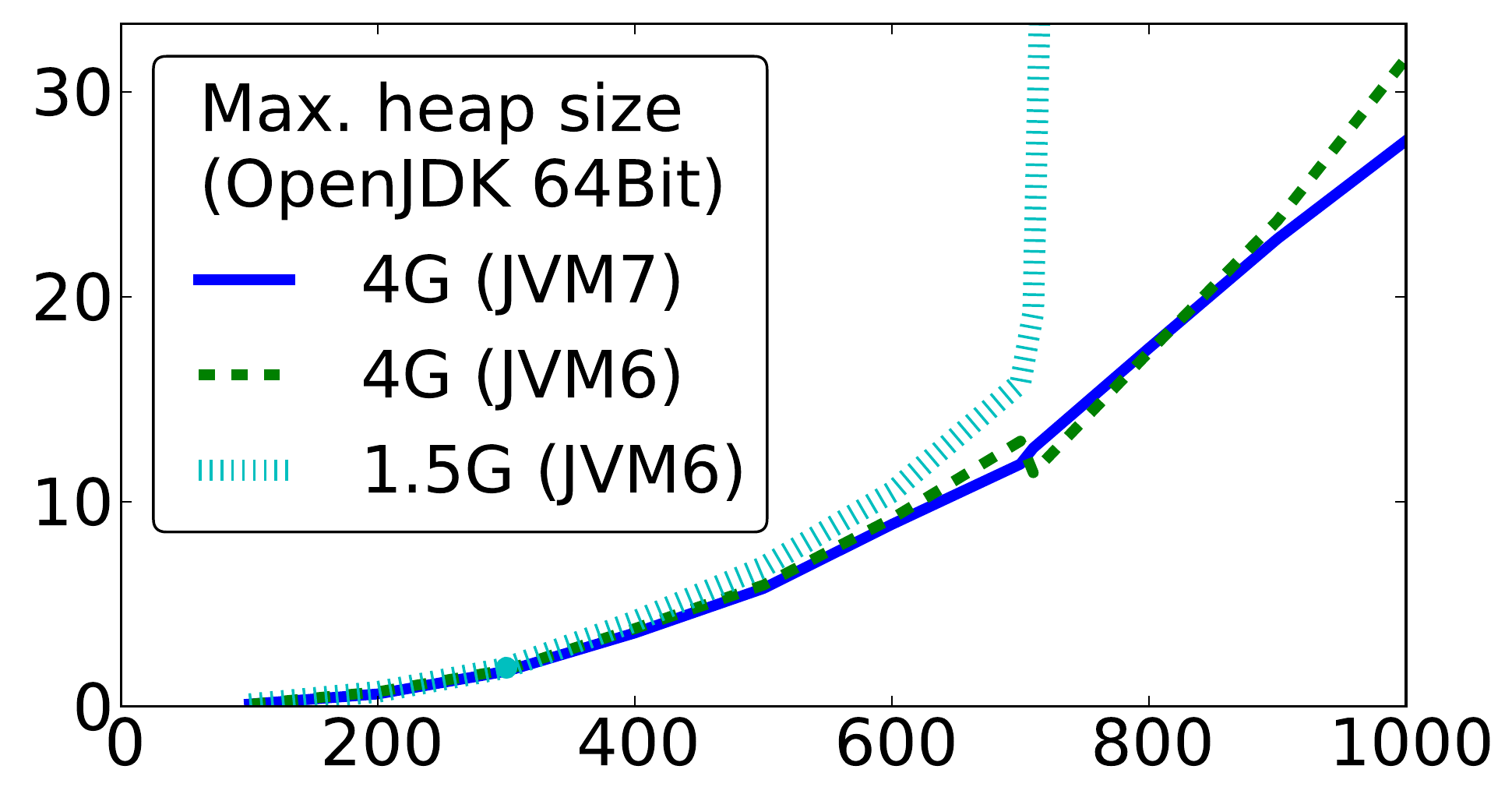}%
}{%
  \caption{Runtime of the policy construction algorithm for $100$ $\Phi$-structured invariants on an i7-2620M CPU (2.70GHz), Java Virtual Machine. X-axis: $\vert V \vert$, Y-axis: runtime in minutes.}
  \label{fig:benchmark}%
}%
\capbtabbox{%
\begin{tabular}{ @{} l @{\hspace*{1em}} l @{\hspace*{1em}} l@{}}%
	\multicolumn{3}{ @{} c @{}}{Section~\ref{sec:case-study} User Case Study: Statistics}\\
	\toprule
	Valid               & Violations                & Missing                   \\
	\midrule
	\studyExperTopoValid & \studyExperTopoInvalid & \studyExperTopoMissing   \\
	\studyInterTopoValid & \studyInterTopoInvalid & \studyInterTopoMissing   \\
	\studyNovicTopoValid & \studyNovicTopoInvalid & \studyNovicTopoMissing   \\
	\midrule
	\multicolumn{3}{@{}l@{}}{\footnotesize{median/arithmetic mean/std deviation}}\\ 
 \bottomrule%
\end{tabular}%
}{%
  \caption{Statistics on Section~\ref{sec:case-study}'s user-designed policies. Number of valid, violating, and missing flows. User experiences (top to bottom): Expert, Intermediate, Novice. Five participants each. } 
  \label{tab:userstudy}%
}%
\end{floatrow}%
\vskip-20pt
\end{figure}

\smallskip
\noindent
\textbf{Computational Complexity.\ } 
\tool{} performs linear in the number of security invariants and quadratic in the number of hosts for $\Phi$-structured invariants. 
For scenarios with less than $100$ hosts, it responds interactively in less than $10$ seconds. 
A benchmark of the automated policy construction, the most expensive algorithm, is presented in Fig.~\ref{fig:benchmark}. 
For $\vert V \vert$ hosts, $\vert V \vert^2 / 4$ flows were created. 
With reasonable memory consumption, policies with up to $250$k flows can be processed in less than half an hour. 
\tool{} contains a lot of machine generated code that is not optimized for performance but correctness. 
However, the overall theoretical and practical performance is sufficient for real-world usage. 
During our work with Airbus Group, we never encountered any performance issues. 

\section{Case Study: A Cabin Data Network}
\label{sec:case-study}
In this section, we present a slightly more complex scenario: a policy for a cabin data network for the general civil aviation. 
This example was chosen as security is very important in this domain and it provides a challenging interaction of different security invariants. 
It is a small imaginary toy example, developed in collaboration with Airbus Group. 
To make it self-contained and accessible to readers without aeronautical background knowledge, it does not obey aeronautical standards. 
However, the scenario is plausible, \ie a real-world scenario may be similar. 
During our research, we also evaluated real world scenarios in this domain. 
With this experience, we try to present a small, simplified, self-contained, plausible toy scenario that, however, preserves many real world snares.

To estimate the scenario's complexity, we asked \studyTotalEntries{} network professionals to design its policy. 
On the one hand, as many use cases as possible should be fulfilled, on the other hand, no security violation must occur. 
Therefore, the task was to maximize the allowed flows without violating any security invariant. 
The results are illustrated in Table~\ref{tab:userstudy}. 
Surprisingly, even expert network administrators made errors (both missing flows and security violations) when designing the policy. 

A detailed scenario description, the host attribute mappings, and raw data are available in~\cite{userstudycabinnetwork2013}. 
Using this reference, we also encourage the active reader to design the policy by oneself before it is revealed in Fig.~\ref{fig:cabinnetwork}. 
The scenario is presented in the following compressed two paragraphs.

\medskip

\begin{footnotesize}

\noindent The network consists of the following hosts. \\
\noindent
\begin{minipage}[t]{0.49\textwidth}
\begin{description}[\compact\setlabelphantom{}]
	\item[CC] The Cabin Core Server, a server that controls essential aircraft features, such as air conditioning and the wireless and wired telecommunication of the crew.
	\item[C1, C2] Two mobile devices for the crew to help them organize, \eg communicate, make announcements.
	\item[Wifi]	A wifi hotspot that allows passengers to access the Internet with their own devices. Explicitly listed as it might also be responsible for billing passenger's Internet access.
\end{description}
\end{minipage}
\hspace*{\fill}
\begin{minipage}[t]{0.49\textwidth}
\begin{description}[\compact\setlabelphantom{}]
	\item[IFEsrv] The In-Flight Entertainment server with movies, Internet access, etc. Master of the IFE displays.
	\item[IFE1, IFE2] Two In-Flight Entertainment displays, mounted at the back of passenger seats. They provide movies and Internet access. Thin clients, everything is streamed from the IFE server.
	\item[P1, P2] Two passenger-owned devices, \eg laptops, smartphones.
	\item[Sat]	A satellite uplink to the Internet.
\end{description}
\end{minipage}

\begin{figure*}[h!tp]
  \centering
  \hspace*{\fill}%
  \begin{subfigure}[t]{0.64\textwidth}
       \captionsetup{width=0.99\textwidth}
       \centering
  		\includegraphics[width=0.99\linewidth]{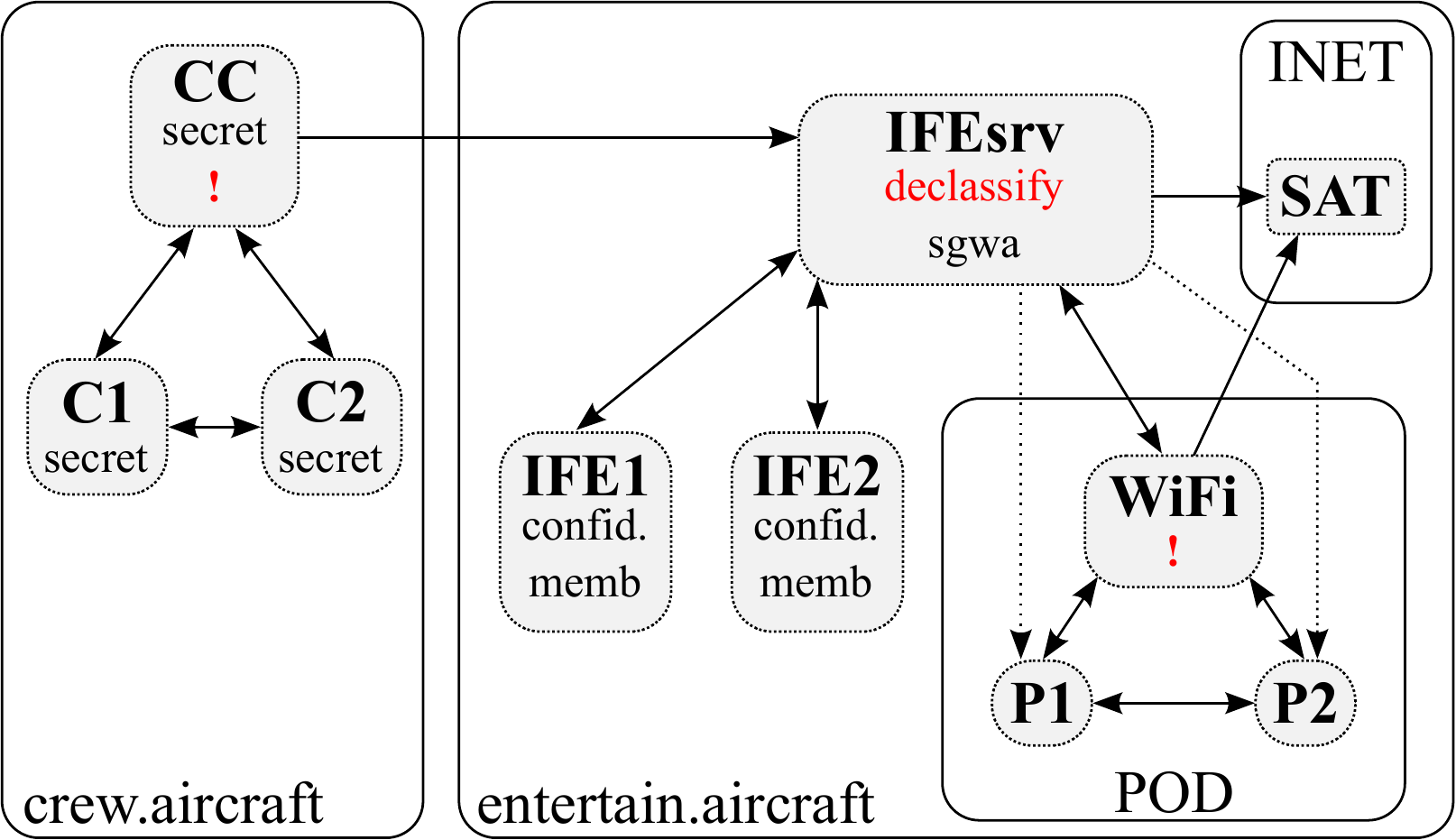}
  \end{subfigure}%
  \hspace*{\fill}%
  \begin{subfigure}[b]{0.34\textwidth}
       \captionsetup{width=0.85\textwidth,labelformat=cornyTop}
       \centering
  		\includegraphics[width=0.70\linewidth]{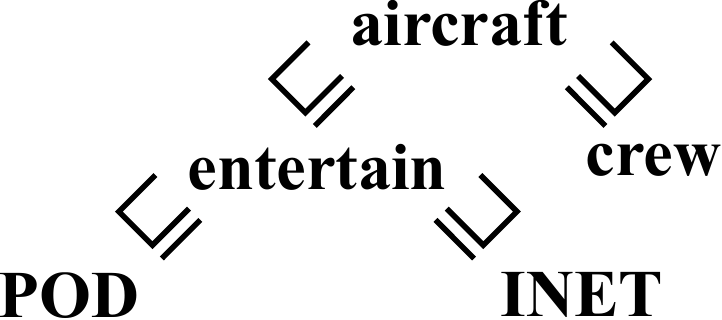}
  		\vspace*{0.5em}
		\caption{Security domains of the cabin data network.}
		\label{fig:cabinnetworkhierarchy}
       \captionsetup{width=0.85\textwidth,labelformat=cornyLeft}
  		\caption{Cabin network policy and hosts' attributes.}
  		\label{fig:cabinnetwork}
  \end{subfigure}%
  \hspace*{\fill}%
\end{figure*}

\noindent
The following three security invariants are specified.%
\begin{description}[\compact\setlabelphantom{}]%
\item[Security Invariant 1, Domain Hierarchy.] Four different security domains exist in the aircraft, c.f. Fig.~\ref{fig:cabinnetworkhierarchy}. 
They separate the \textbf{crew} domain, the \textbf{entertain}ment domain, the passenger-owned devices (\textbf{POD}) domain and the Internet (\textbf{INET}) domain. 
In Fig.~\ref{fig:cabinnetwork}, the host domain mapping is illustrated and trusted devices are marked with an exclamation mark. 

The CC may send to the entertain domain, hence it is trusted. 
Possible use cases: Stewards coordinate food distribution; Announcement from the crew is send to the In-Flight Entertainment system (via CC) and distributed there to the IFE displays.

The Wifi is located in the \textbf{POD} domain to be reachable by PODs. It is trusted to send to the entertain domain. 
Possible use case: Passenger subscribes a film from the IFE server to her notebook or establishes connections to the Internet. 

In the \textbf{INET} domain, the SAT is isolated to prevent accesses from the Internet into the aircraft. 

\item[Security Invariant 2, Security Gateway.]
The IFE displays are thin clients and strictly bound to their server. Peer to peer communication is prohibited. 
The Security Gateway model directly provides the respective access control restrictions.

\item[Security Invariant 3, Bell LaPadula with Trust.]
Invariant 3 defines information flow restrictions by labeling confidential information sources. 
To protect the passenger's privacy when using the IFE displays, it is undesirable that the IFE displays communicate with anyone, except for the IFEsrv. 
Therefore, the IFE displays are marked as confidential (confid.).
The IFEsrv is considered a central trusted device. 
To enable passengers to surf the Internet on the IFE displays by forwarding the packets to the Internet or forward announcements from the crew, it must be allowed to declassify any information to the default (\ie unclassified) security clearance. 
Finally, the crew communication is considered more critical than the convenience features, therefore, CC, C1, and C2 are considered secret. 
As the IFEsrv is trusted, it can receive and forward announcements from the crew. 
\end{description}
\end{footnotesize}%
\noindent This case study illustrates that this complex scenario can be divided into three security invariants that can be represented with the help of the previously presented templates. 
Figure~\ref{fig:cabinnetwork} also reveals that very few host attributes must be manually specified; the automatically added secure default attributes are not shown. 
All security invariants are fulfilled.

The automated policy construction yields the following results. 
The solid edges unified with the dashed edges\footnote{unified with all reflexive edges, \ie in-host communication} result in the uniquely defined policy with the maximum number of allowed flows. 
The solid lines were given by the policy, the dashed lines were calculated from the invariants. 
These `diffs' are computed and visualized automatically by \tool{}. 
They provide the end user with helpful feedback regarding `\emph{what do my invariants require?}' vs. `\emph{what does my policy specify?}'.
This results in a feedback loop we used extensively during our research to refine the policy and the invariants. 
It provides a `feeling' for the invariants.

The main evaluation of this work are the formal correctness proofs. 
However, we also presented \tool{} to the \studyTotalEntries{} users and asked them to personally judge \tool{}'s utility. 
It was considered downright helpful and a majority would want to use it for similar tasks. 
The graphical feedback was also much appreciated.

\section{Related Work}
\label{sec:relatedwork}
In a field study with 38 participants, Hamed and Al-Shaer discovered that ``even expert administrators can make serious mistakes when configuring the network security policy''~\cite{netsecconflicts}. 
Our user feedback session extends this finding as we discovered that even expert administrators can make serious mistakes when \emph{designing} the network security policy. 

In their inspiring work, Guttman and Herzog~\cite{guttman05rigorous} describe a formal modeling approach for network security management. 
They suggest algorithms to verify whether configurations of firewalls and IPsec gateways fulfill certain security goals. 
These comparatively low-level security goals may state that a certain packet's path only passes certain areas or that packets between two networked hosts are protected by IPsec's ESP confidentiality header. 
This allows reasoning on a lower abstraction level at the cost of higher manual specification and configuration effort. 
Header space analysis~\cite{kazemian2012HSA} allows checking static network invariants such as no-forwarding-loops or traffic-isolation on the forwarding and middleboxes plane. 
It provides a common, protocol-agnostic framework and algebra on the packet header bits. 

Firmato~\cite{bartal1999firmato} was designed to ease management of firewalls. 
A firewall-in\-de\-pen\-dent entity relationship model is used to specify the security policy.
With the help of a model compiler, such a model can be translated to firewall configurations. 
Ethane~\cite{ethane07} is a link layer security architecture which evolved to the network operating system NOX~\cite{gude2008nox}. 
They implement high-level security policies and propose a secure binding from host names to network addresses. 
In the long term, we consider \tool{} a valuable add-on on top of such systems for policy verification. 
For example, it could warn the administrator that a recent network policy change violates a security invariant, maybe defined years ago.

Expressive policy specification languages, such as Ponder~\cite{ponder2001}, were proposed. 
Positive authorization policies (only a small aspect of Ponder) are roughly comparable to our policy graph. 
The authors note that \eg negative authorization policies (deny-rules) can create conflicts. 
Policy constraints can be checked at compile time.
In~\cite{policy2010berapolicyformalenterprise}, a policy specification language (SPSL) with allow and deny policy rules is presented. 
With this, a conflict-free policy specification is constructed. 
Conflict-free Boolean formulas of this policy description and the policy implementation in the security mechanisms (router ACL entries) are checked for equality using a SAT solver.
One unique feature covered is hidden service access paths, \eg http might be prohibited in zone1 but zone1 can ssh to zone2 where http is allowed. 
\cite{policy2009expressivedynamic} focuses on policies in dynamic systems and their analysis. 
These papers require specification of the verification goals and security goals and can thus benefit from our contributions. 

This work's modeling concept is very similar to the Attribute Based Access Control (ABAC) model~\cite{abac2005}, though the underlying formal objects differ. 
ABAC distinguishes subjects, resources, and environments. 
Attributes may be assigned to each of these entities, similar to our host mappings. 
The ABAC policy model consists of positive rules which grant access based on the assigned attributes, comparably to security invariant templates. 
Therefore, our insights and contributions are also applicable to the ABAC model.

\section{Conclusion}
\label{sec:conclusion}
After more than 50k changed lines of formal theory, our simple, yet powerful, model landscape emerged. 
Representing policies as graphs makes them visualizable. 
Describing security invariants as total Boolean-valued functions is both expressive and accessible to formal analysis. 
Representing host mappings as partial configurations is end-user-friendly, transforming them to total functions makes them handy for the design of templates. 
With this simple model, we discovered important universal insights on security invariants. 
In particular, the transformation of host mappings and a simple sanity check which guarantees that security policy violations can always be resolved. 
This provides deep insights about how to express verification goals. 
The full formalization in the Isabelle/HOL theorem prover provides high confidence in the correctness.

\subsubsection*{Acknowledgments \& Availability.}
\addcontentsline{toc}{section}{Acknowledgment}
We thank all the participants of our feedback session and the anonymous reviewers very much. 
A special thanks goes to our colleague Lothar Braun for his valuable feedback. 
Lars Hupel helped finalizing the paper. 
This work has been supported by the German Federal Ministry of Education and Research (BMBF) under support code 16BY1209F, project ANSII, 
and 16BP12304, EUREKA project SASER, and by the European Commission under the FP7 project EINS, grant number 288021.

\medskip 
\noindent
Our Isabelle/HOL theory files and \tool{}'s Scala source code are available at 
\begin{center}%
\vskip-5pt
\url{https://github.com/diekmann/topoS}%
\end{center}%
\bibliographystyle{splncs03}
\bibliography{literature_arxiv}
\parindent 0pt%
\begin{minipage}{0.9999\textwidth}%
\begingroup%
\def\enoteformat{\rightskip=0pt \leftskip=0pt \parindent=0pt \leavevmode{\makeenmark}}%
\def\enotesize{\footnotesize}%
\begin{footnotesize} 
\theendnotes%
\end{footnotesize}%
\endgroup%
\end{minipage}%
%
%

\clearpage\newpage
\begin{appendix}
\section*{Appendix}
In this appendix, we provide the host attribute mappings of Section~\ref{sec:case-study}'s case study. 

\medskip

Domain Hierarchy
\begin{IEEEeqnarray*}{l "c" l}
\mathrm{CC} & \mapsto & \left(\mathrm{level}:\ \mathit{crew}.\mathit{aircraft},\ \ \mathrm{trust}:\ 1\right)\\
\mathrm{C1} & \mapsto & \left(\mathrm{level}:\ \mathit{crew}.\mathit{aircraft},\ \ \mathrm{trust}:\ 0\right)\\
\mathrm{C2} & \mapsto & \left(\mathrm{level}:\ \mathit{crew}.\mathit{aircraft},\ \ \mathrm{trust}:\ 0\right)\\
\mathrm{IFEsrv} & \mapsto & \left(\mathrm{level}:\ \mathit{entertain}.\mathit{aircraft},\ \ \mathrm{trust}:\ 0\right)\\
\mathrm{IFE1} & \mapsto & \left(\mathrm{level}:\ \mathit{entertain}.\mathit{aircraft},\ \ \mathrm{trust}:\ 0\right)\\
\mathrm{IFE2} & \mapsto & \left(\mathrm{level}:\ \mathit{entertain}.\mathit{aircraft},\ \ \mathrm{trust}:\ 0\right)\\
\mathrm{SAT} & \mapsto & \left(\mathrm{level}:\ \mathit{INET}.\mathit{entertain}.\mathit{aircraft},\ \ \mathrm{trust}:\ 0\right)\\
\mathrm{Wifi} & \mapsto & \left(\mathrm{level}:\ \mathit{POD}.\mathit{entertain}.\mathit{aircraft},\ \ \mathrm{trust}:\ 1\right)\\
\mathrm{P1} & \mapsto & \left(\mathrm{level}:\ \mathit{POD}.\mathit{entertain}.\mathit{aircraft},\ \ \mathrm{trust}:\ 0\right)\\
\mathrm{P2} & \mapsto & \left(\mathrm{level}:\ \mathit{POD}.\mathit{entertain}.\mathit{aircraft},\ \ \mathrm{trust}:\ 0\right)\\
\end{IEEEeqnarray*}

Security Gateway
\begin{IEEEeqnarray*}{l "c" l}
\mathrm{IFEsrv} & \mapsto & \mathit{sgwa}\\
\mathrm{IFE1} & \mapsto & \mathit{memb}\\
\mathrm{IFE2} & \mapsto & \mathit{memb}\\
\end{IEEEeqnarray*}

Simplified Bell LaPadula with Trust
\begin{IEEEeqnarray*}{l "c" l}
\mathrm{CC} & \mapsto & \left(\mathrm{sc}:\ \mathit{secret},\ \ \mathrm{trust}:\ \mathit{False}\right)\\
\mathrm{C1} & \mapsto & \left(\mathrm{sc}:\ \mathit{secret},\ \ \mathrm{trust}:\ \mathit{False}\right)\\
\mathrm{C2} & \mapsto & \left(\mathrm{sc}:\ \mathit{secret}\,\ \ \mathrm{trust}:\ \mathit{False}\right)\\
\mathrm{IFE1} & \mapsto & \left(\mathrm{sc}:\ \mathit{confidential},\ \ \mathrm{trust}:\ \mathit{False}\right)\\
\mathrm{IFE2} & \mapsto & \left(\mathrm{sc}:\ \mathit{confidential},\ \ \mathrm{trust}:\ \mathit{False}\right)\\
\mathrm{IFEsrv} & \mapsto & \left(\mathrm{sc}:\ \mathit{unclassified},\ \ \mathrm{trust}:\ \mathit{True}\right)\\
\end{IEEEeqnarray*}

\end{appendix}
\end{document}